\documentclass[runningheads, envcountsame, a4paper]{llncs}
    \usepackage[utf8]{inputenc}
    \usepackage{booktabs}
    \usepackage{amsmath}
    \usepackage{graphicx}
    \usepackage{subfig}
    \usepackage{tikz}
    \usetikzlibrary{positioning,backgrounds,arrows.meta}
    \usepackage{pgfplots}
    \pgfplotsset{compat=1.6}
\begin{document}
    \title{A Comparison of Adaptation Techniques and Recurrent Neural Network Architectures}
    \titlerunning{A Comparison of Adaptation Techniques and RNN Architectures}
    %
    \author{Jan Van\v{e}k\orcidID{0000-0002-2639-6731} \and
    Josef Mich\'{a}lek\orcidID{0000-0001-7757-3163} \and \\
    Jan Zelinka\orcidID{0000-0001-6834-9178} \and
    Josef Psutka\orcidID{0000-0002-0764-3207}}
    \authorrunning{J. Van\v{e}k, J. Mich\'{a}lek, J. Zelinka, J. Psutka}
    %
    \institute{University of West Bohemia \\ Univerzitn\'{i} 8, 301 00 Pilsen, Czech Republic\\
    \email{\{vanekyj,orcus,zelinka,psutka\}@kky.zcu.cz}}
    \maketitle              
    \begin{abstract}
Recently, recurrent neural networks have become state-of-the-art in acoustic modeling for automatic speech recognition. The long short-term memory (LSTM) units are the most popular ones. However, alternative units like gated recurrent unit (GRU) and its modifications outperformed LSTM in some publications. In this paper, we compared five neural network (NN) architectures with various adaptation and feature normalization techniques. We have evaluated feature-space maximum likelihood linear regression, five variants of i-vector adaptation and two variants of cepstral mean normalization. The most adaptation and normalization techniques were developed for feed-forward NNs and, according to results in this paper, not all of them worked also with RNNs. For experiments, we have chosen a well known and available TIMIT phone recognition task. The phone recognition is much more sensitive to the quality of AM than large vocabulary task with a complex language model. Also, we published the open-source scripts to easily replicate the results and to help continue the development.

    \end{abstract}
    \section{Introduction}
    Neural Networks (NNs) and deep NNs (DNNs) became dominant in the field of the acoustic modeling several years ago. Simple feed-forward (FF) DNNs were faded away in recent years. The current progress is based on the modeling of a longer temporal context of individual feature frames. Main two ways are actually popular: First, a larger context is modeled by a time-delayed NN (TDNN) \cite{Waibel1989}, \cite{Peddinti2015}. TDNNs model long term temporal dependencies with training times comparable to standard feed-forward DNNs. In the TDNN architecture, the initial transforms learn narrow contexts and the deeper layers process the hidden activations from a wider temporal context. Hence the higher layers have the ability to learn wider temporal relationships. The second way to learn the longer temporal context is to use recurrent NNs (RNNs). The most popular RNN architecture is a long short-term memory (LSTM) that has been designed to address the vanishing and exploding gradient problems of conventional RNNs. Unlike feed-forward neural networks, RNNs have cyclic connections making them powerful for modeling sequences \cite{Sak2014}. The main drawback is much slower training due to the sequential nature of the learning algorithm. An unfolding of the recurrent network during training was proposed in \cite{Saon2014} to speed-up the training, however it is still significantly slower than FF NNs or TDNNs. More recently, another type of recurrent unit, a gated recurrent unit (GRU), was proposed in \cite{Cho2014}, \cite{chung2014empirical}. Similarly to the LSTM unit, the GRU has gating units that modulate the flow of information inside the unit, however, without having a separate memory cells. Further revising GRUs leaded to a simplified architecture potentially more suitable for speech recognition in \cite{Ravanelli_IS2017}. First, removing the reset gate in the GRU design resulted in a simpler single-gate architecture called modified GRU (M-GRU). Second, replacing \emph{tanh} with \emph{ReLU} activations in the state update equations was proposed and called M-reluGRU. A more detailed overview of the RNN architectures follows in Section \ref{RNN}.

Even if large datasets are used for the DNN training, an adaptation of an acoustic model (AM) to a test speaker and environment is beneficial. A lot of techniques have been reported on the adaptation, such as the classical maximum a posterior (MAP) and maximum likelihood linear regression (MLLR) for traditional GMM-HMM acoustic models. Although this technique can be modified for an NN-based acoustic model, a much simpler application has so-called feature space MLLR (fMLLR)~ \cite{Gales199875} because fMLLR changes only features and it does not adapt NN parameters. This speaker adaptation technique can be easily applied in an NN-based acoustic model \cite{Seide2011}, \cite{ParthasarathiHM15}, \cite{RathPVC13}. Therefore, fMLLR can be used to any DNN architecture. I-vectors originally developed for speaker recognition can be used to the speaker and environment adaptation also \cite{Saon2013}, \cite{Karafiat2011}. Alternative approach is using of discriminative speaker codes \cite{Xue2014}, \cite{Huang2016}. More detailed description of the adaptation techniques used in this paper follows.

    \section{Adaptation of DNNs}
		\label{adaptation}
    The simplest way of the adaptation is a feature level adaptation. When adapting input features, NN can have any structure. The most popular and well known technique is fMLLR based on an underlying HMM-GMM that is used during initial stage of the NN training.
\subsection{fMLLR}
The fMLLR transforms feature frames with a speaker-specific square matrix $A$ and a bias $b$. For HMM-GMMs, $A$ and $b$ are estimated to maximize the likelihood of the adaptation data given the model \cite{Gales199875}, \cite{Seide2011}. In the training phase, the speaker-specific transform may be updated several-times alternating HMM-GMM update. These approach is usually called a speaker adaptive training (SAT). The result of the training phase is a canonical model that requires using adaptation during testing phase. However, two-pass processing is required during test phase. The first pass produces unsupervised alignment that is used to estimate the transform parameters via maximum likelihood. The model used for alignment does not need to be the identical model to the final canonical one. Because all the steps are using the underline HMM-GMM any NN architecture may be used to train the final NN acoustic model.
\subsection {i-vectors}
The i-vector extraction is a well known technique, so we focused here to more practical points. An detailed description of i-vectors can be found in \cite{Saon2013}, \cite{Karafiat2011} and further papers referenced in there.

The i-vector extraction is comprised from following steps:

\begin{enumerate}
\item An universal background model (UBM) needs to be trained. Usually a GMM with 512 to 2048 diagonal components is used. The quality of GMM is not critical, so some speed-up methods can be utilized. Features for UBM do not need to match witch features for NN nor i-vector accumulators. Usually, features with cepstral mean normalization (CMN) or cepstral mean and variance normalization (CMVN) are used for UBM. The normalization techniques reduce speaker and environment variability. Features without any normalization are used for the i-vector accumulators to carry more speaker- and environment-related information.
\item Zero-order and centered first-order statistics are accumulated for every speaker according to the UBM posteriors.
\item The i-vector extraction transforms are estimated iteratively by expectation/maximization (EM) algorithm.
\item The i-vector for individual speakers is evaluated. For training speakers, zero-order and centered first-order statistics have been already accumulated. For other speakers, statistics must be accumulated. Then, the i-vector is evaluated by the i-vector extraction transforms computed in the third step.
\end{enumerate}
The four step process seems simple but there are some details that need to be mentioned:
\begin{itemize}
\item CMN or CMVN may be computed online or offline. The offline variant may be per-utterance or per-speaker. The online variant starts from global cepstral mean and it is subsequently updated. An exponential forgetting is usable for very long utterances. The training setup should match with the testing one.
\item The accumulated statistic should be saturated or scaled-down for long utterances due to an i-vector overfitting.
\item The offline scenario is not proper for training. The number of speakers and thus variants of i-vectors is very limited and leads to NN overfitting. The online scenario is recommended for training, in the Kaldi Switchboard example recipe the number of speakers is also boosted by pseudo-speakers. Two training utterances represent one pseudo-speaker. The offline scenario may be used in the test phase.
\end{itemize}

    \section{Recurrent Neural Network Architectures}
		\label{RNN}
    \subsection{Long Short-Term Memory}
\label{sec:lstm}
Long short-term memory (LSTM) is a widely used type of recurrent neural network (RNN).
Standard RNNs suffer from both exploding and vanishing gradient problems.
Both of these problems are caused by the fact, that information flowing through the RNN passes through many stages of multiplication.
The gradient is essentially equal to the weight matrix raised to a high power.
This results in the gradient growing or shrinking at an exponential rate to the number of timesteps.

The exploding gradient problem can be solved simply by truncating the gradient.
On the other hand, the vanishing gradient problem is harder to overcome.
It does not simply cause the gradient to be small; the gradient components corresponding to long-term dependencies are small while the components corresponding to short-term dependencies are large.
Resulting RNN can then learn short-term dependencies but not long-term dependencies.

The LSTM was proposed in 1997 by Hochreiter and Scmidhuber \cite{hochreiter1997long} as a solution to the vanishing gradient problem.
Let $c_t$ denote a hidden state of a LSTM.
The main idea is that instead of computing $c_t$ directly from $c_{t-1}$ with matrix-vector product followed by an activation function, the LSTM computes $\Delta c_t$ and adds it to $c_{t-1}$ to get $c_t$.
The addition operation is what eliminates the vanishing gradient problem.

Each LSTM cell is composed of smaller units called gates, which control the flow of information through the cell.
The forget gate $f_t$ controls what information will be discarded from the cell state, input gate $i_t$ controls what new information will be stored in the cell state and output gate $o_t$ controls what information from the cell state will be used in the output.

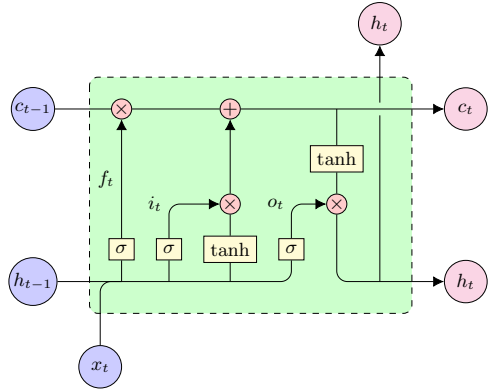
\begin{figure}
    \centering
    \scalebox{0.8}{
        \begin{tikzpicture}[node distance=10pt]
    \tikzstyle{phantom}=[inner sep=0,outer sep=0,minimum size=0]
    \tikzstyle{pointwise}=[draw,circle,fill=red!20!white,inner sep=0]
    \tikzstyle{layer}=[draw,rectangle,fill=yellow!20!white]
    \tikzstyle{transition}=[-{Latex[length=5pt,width=5pt]},rounded corners=5]
    \tikzstyle{line}=[rounded corners=5]
    \tikzstyle{node_in}=[draw,circle,fill=blue!20!white,inner sep=1pt,minimum size=20pt]
    \tikzstyle{node_out}=[draw,circle,fill=magenta!20!white,inner sep=1pt,minimum size=20pt]

    \node[layer] (forget_gate) {$\sigma$};
    \node[layer,right=of forget_gate] (input_gate) {$\sigma$};
    \node[layer,right=of input_gate] (input_tanh) {$\tanh$};
    \node[layer,right=of input_tanh] (output_gate) {$\sigma$};
    \node[pointwise,above=of input_tanh] (input_mult) {$\times$};
    \node[pointwise,above=35pt of input_mult] (input_add) {$+$};
    \node[pointwise,right=40pt of input_mult] (output_mult) {$\times$};
    \node[layer,above=of output_mult] (output_tanh) {$\tanh$};

    \node[phantom,below=of forget_gate] (forget_below) {};
    \node[phantom] at(forget_below -| input_gate) (input_below) {};
    \node[phantom] at(forget_below -| input_tanh) (input_tanh_below) {};
    \node[node_in,left=30pt of forget_below] (h_in) {$h_{t-1}$};
    \node[phantom,right=70pt of input_tanh_below] (output_below) {};
    \node[node_out,right=30pt of output_below] (h_out) {$h_t$};
    \node[node_in,below=30pt of forget_below,xshift=-10pt] (input_x) {$x_t$};

    \node[pointwise] at (forget_gate |- input_add) (forget_mult) {$\times$};
    \node[node_in] at(forget_mult -| h_in) (c_in) {$c_{t-1}$};
    \node[phantom] at(forget_mult -| output_tanh) (output_tanh_above) {};
    \node[node_out] at(forget_mult -| h_out) (c_out) {$c_t$};
    \node[phantom] at(forget_mult -| output_below) (output_above) {};
    \node[node_out,above=30pt of output_above] (output_h) {$h_t$};

    \draw[transition] (forget_gate) -- (forget_mult) node[midway,left] {$f_t$};
    \draw[transition] (input_gate) |- (input_mult) node[midway,left] {$i_t$};
    \draw[line] (input_tanh) -- (input_mult);
    \draw[transition] (input_mult) -- (input_add);
    \draw[transition] (output_gate) |- (output_mult) node[midway,left] {$o_t$};
    \draw[line] (output_tanh_above) -- (output_tanh);
    \draw[line] (output_tanh) -- (output_mult);

    \draw[line] (h_in) -| (output_gate);
    \draw[line] (forget_below) -- (forget_gate);
    \draw[line] (input_below) -- (input_gate);
    \draw[line] (input_tanh_below) -- (input_tanh);
    \draw[transition] (output_mult) |- (h_out);
    \draw[line] (input_x) |- (forget_below);

    \draw[line] (c_in) -- (forget_mult);
    \draw[line] (forget_mult) -- (input_add);
    \draw[transition] (input_add) -- (c_out);
    \draw[line,shorten >=3pt] (output_below) -- (output_above);
    \draw[transition,shorten <=3pt] (output_above) -- (output_h);

    \begin{scope}[on background layer]
        \path (forget_mult)+(-15pt,15pt) node (a) {};
        \path (output_below)+(15pt,-15pt) node (b) {};
        \path[draw,dashed,fill=green!20!white,rounded corners] (a) rectangle (b);
    \end{scope}
\end{tikzpicture}
    }
    \caption{Structure of a LSTM unit}
    \label{fig:lstm}
\end{figure}

The LSTM has two hidden states, $c_t$ and $h_t$.
The state $c_t$ fights the gradient vanishing problem while $h_t$ allows the network to make complex decisions over short periods of time.
There are several slightly different LSTM variants.
The architecture used in this paper is specified by the following equations:
\begin{align*}
    i_t &= \sigma(W_{xi} x_t + W_{hi} h_{t-1} + b_i) \\
    f_t &= \sigma(W_{xf} x_t + W_{hf} h_{t-1} + b_f) \\
    o_t &= \sigma(W_{xo} x_t + W_{ho} h_{t-1} + b_o) \\
    c_t &= f_t \ast c_{t-1} + i_t \ast \tanh(W_{xc} x_t + W_{hc} h_{t-1} + b_c) \\
    h_t &= o_t \ast \tanh(c_t)
\end{align*}
The figure \ref{fig:lstm} shows the internal structure of LSTM.

\subsection{Gated Recurrent Unit}
A gated recurrent unit (GRU) was proposed in 2014 by Cho et al.\cite{chung2014empirical}
Similarly to the LSTM unit, the GRU has gating units that modulate the flow of information inside the unit, however, without having a separate memory cells.

The update gate $z_t$ decides how much the unit updates its activation and reset gate $r_t$ determines which information will be kept from the old state.
GRU does not have any mechanism to control what information to output, therefore it exposes the whole state.

The main differences between LSTM unit and GRU are:
\begin{itemize}
    \item GRU has 2 gates, LSTM has 3 gates
    \item GRUs do not have an internal memory different from the unit output, LSTMs have an internal memory $c_t$ and the output is controlled by an output gate
    \item Second nonlinearity is not applied when computing the output of GRUs
\end{itemize}

The GRU unit used in this work is described by the following equations:
\begin{align*}
    r_t &= \sigma(W_r x_t + U_r h_{t - 1} + b_r) \\
    z_t &= \sigma(W_z x_t + U_z h_{t - 1} + b_z) \\
    \tilde{h_t} &= \tanh(W x_t + U (r_t \ast h_{t - 1}) + b_h) \\
    h_t &= (1 - z_t) \ast h_{t-1} + z_t \ast \tilde{h_t} \\
\end{align*}

\subsection{Modified Gated Recurrent Unit with ReLU}
Ravanelli introduced a simplified GRU architecture, called \emph{M-reluGRU}, in \cite{Ravanelli_IS2017}.
This simplified architecture does not have the reset gate and uses ReLU as an activation function instead of tanh.

The \emph{M-reluGRU} unit is described by the following equations:
\begin{align*}
    z_t &= \sigma(W_z x_t + U_z h_{t - 1} + b_z) \\
    \tilde{h_t} &= \text{ReLU}(W x_t + U h_{t - 1} + b_h) \\
    h_t &= (1 - z_t) \ast h_{t-1} + z_t \ast \tilde{h_t} \\
\end{align*}

We have also used this unit with the reset gate to evaluate the impact of the missing reset gate on the network performance.
This unit is effectivelly a normal GRU with ReLU as an activation function and we called it \emph{reluGRU} in this paper.

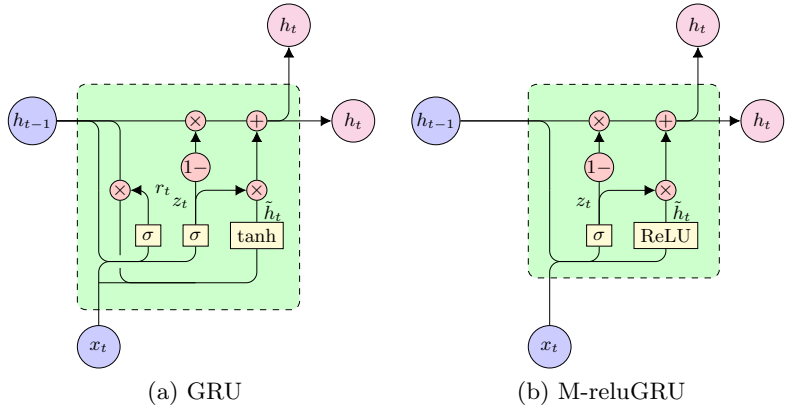
\begin{figure}
    \centering
    \subfloat[GRU]{
    \scalebox{0.8}{
        \begin{tikzpicture}[node distance=10pt]
    \tikzstyle{phantom}=[inner sep=0,outer sep=0,minimum size=0]
    \tikzstyle{pointwise}=[draw,circle,fill=red!20!white,inner sep=0]
    \tikzstyle{layer}=[draw,rectangle,fill=yellow!20!white]
    \tikzstyle{transition}=[-{Latex[length=5pt,width=5pt]},rounded corners=5]
    \tikzstyle{line}=[rounded corners=5]
    \tikzstyle{node_in}=[draw,circle,fill=blue!20!white,inner sep=1pt,minimum size=20pt]
    \tikzstyle{node_out}=[draw,circle,fill=magenta!20!white,inner sep=1pt,minimum size=20pt]

    \node[layer] (reset_gate) {$\sigma$};
    \node[layer,right=of reset_gate] (update_gate) {$\sigma$};
    \node[layer,right=of update_gate] (tanh) {$\tanh$};
    \node[pointwise,above=of tanh] (tanh_mult) {$\times$};
    \node[pointwise,above=20pt of update_gate] (minus_one) {$1-$};
    \node[pointwise,above=of minus_one] (minus_mult) {$\times$};
    \node[pointwise] at(minus_mult -| tanh_mult) (tanh_add) {$+$};

    \node[phantom,below left=of reset_gate] (reset_below) {};
    \node[phantom,left=of reset_below] (reset_belowl) {};
    \node[phantom,below=of reset_below] (reset_belowd) {};
    \node[phantom] at(reset_belowl |- reset_belowd) (reset_belowld) {};
    \node[phantom] at(update_gate |- reset_belowd) (reset_belowrd) {};

    \node[pointwise] at(tanh_mult -| reset_below) (reset_mult) {$\times$};
    \node[node_in,below=20pt of reset_belowld] (input_x) {$x_t$};
    \node[node_out,right=30pt of tanh_add] (h_out) {$h_t$};
    \node[node_out,above=30pt of tanh_add,xshift=15pt] (output_h) {$h_t$};

    \node[phantom] at(reset_mult -| reset_belowl) (reset_above) {};
    \node[node_in,left=60pt of minus_mult] (h_in) {$h_{t-1}$};

    \draw[transition] (reset_gate) |- (reset_mult) node[midway,right] {$r_t$};
    \draw[line] (update_gate) -- (minus_one) node[midway,left] {$z_t$};
    \draw[line] (tanh) -- (tanh_mult) node[midway,right] {$\tilde{h}_t$};
    \draw[transition] (minus_one) -- (minus_mult);
    \draw[transition] (tanh_mult) -- (tanh_add);
    \draw[transition] (update_gate) |- (tanh_mult);

    \draw[line] (reset_below) -| (reset_gate);
    \draw[line] (reset_below) -| (update_gate);
    \draw[line] (input_x) |- (reset_below);
    \draw[line] (reset_belowld) -| (tanh);
    \draw[line] (reset_above) |- (reset_below);
    \draw[line,shorten >=3pt] (reset_mult) -- (reset_below);
    \draw[line,shorten >=3pt] (reset_belowrd) -| (reset_below);

    \draw[line] (h_in) -| (reset_above);
    \draw[line] (h_in) -| (reset_mult);
    \draw[line] (h_in) -- (minus_mult);
    \draw[line] (minus_mult) -- (tanh_add);
    \draw[transition] (tanh_add) -- (h_out);
    \draw[transition] (tanh_add) -| (output_h);

    \begin{scope}[on background layer]
        \path (reset_above)+(-10pt,50pt) node (a) {};
        \path (tanh)+(20pt,-35pt) node (b) {};
        \path[draw,dashed,fill=green!20!white,rounded corners] (a) rectangle (b);
    \end{scope}
\end{tikzpicture}
        }
    }
    \subfloat[M-reluGRU]{
    \scalebox{0.8}{
        \begin{tikzpicture}[node distance=10pt]
    \tikzstyle{phantom}=[inner sep=0,outer sep=0,minimum size=0]
    \tikzstyle{pointwise}=[draw,circle,fill=red!20!white,inner sep=0]
    \tikzstyle{layer}=[draw,rectangle,fill=yellow!20!white]
    \tikzstyle{transition}=[-{Latex[length=5pt,width=5pt]},rounded corners=5]
    \tikzstyle{line}=[rounded corners=5]
    \tikzstyle{node_in}=[draw,circle,fill=blue!20!white,inner sep=1pt,minimum size=20pt]
    \tikzstyle{node_out}=[draw,circle,fill=magenta!20!white,inner sep=1pt,minimum size=20pt]

    \node[layer] (update_gate) {$\sigma$};
    \node[layer,right=of update_gate] (relu) {ReLU};
    \node[pointwise,above=of relu] (relu_mult) {$\times$};
    \node[pointwise,above=20pt of update_gate] (minus_one) {$1-$};
    \node[pointwise,above=of minus_one] (minus_mult) {$\times$};
    \node[pointwise] at(minus_mult -| relu_mult) (relu_add) {$+$};

    \node[phantom,below left=of update_gate] (update_below) {};
    \node[phantom,left=of update_below] (update_belowl) {};
    \node[phantom,below=of update_below] (update_belowd) {};
    \node[phantom] at(update_belowl |- update_belowd) (update_belowld) {};
    \node[phantom] at(update_gate |- update_belowd) (update_belowrd) {};

    \node[node_in,below=20pt of update_belowld] (input_x) {$x_t$};
    \node[node_out,right=30pt of relu_add] (h_out) {$h_t$};
    \node[node_out,above=30pt of relu_add,xshift=15pt] (output_h) {$h_t$};

    \node[phantom] at(relu_mult -| update_belowl) (reset_above) {};
    \node[node_in,left=60pt of minus_mult] (h_in) {$h_{t-1}$};

    \draw[line] (update_gate) -- (minus_one) node[midway,left] {$z_t$};
    \draw[line] (relu) -- (relu_mult) node[midway,right] {$\tilde{h}_t$};
    \draw[transition] (minus_one) -- (minus_mult);
    \draw[transition] (relu_mult) -- (relu_add);
    \draw[transition] (update_gate) |- (relu_mult);

    \draw[line] (update_below) -| (update_gate);
    \draw[line] (input_x) |- (update_below);
    \draw[line] (update_below) -| (relu);
    \draw[line] (reset_above) |- (update_below);

    \draw[line] (h_in) -| (reset_above);
    \draw[line] (h_in) -- (minus_mult);
    \draw[line] (minus_mult) -- (relu_add);
    \draw[transition] (relu_add) -- (h_out);
    \draw[transition] (relu_add) -| (output_h);

    \begin{scope}[on background layer]
        \path (reset_above)+(-10pt,50pt) node (a) {};
        \path (relu)+(25pt,-20pt) node (b) {};
        \path[draw,dashed,fill=green!20!white,rounded corners] (a) rectangle (b);
    \end{scope}
\end{tikzpicture}
        }
    }
    \caption{Structure of GRU and M-reluGRU units}
    \label{fig:gru}
\end{figure}

The figure \ref{fig:gru} shows the internal structure and difference of GRU and M-reluGRU units.

    \section{Experiments}
		\label{experiments}
    We have chosen TIMIT, a small phone recognition task, as a benchmark of the NN architectures and adaptation techniques. The TIMIT corpus is well known and available. The small size allows a rapid testing and simulates a low-resource scenario that is still an issue for many minor languages. The phone recognition is much more sensitive to quality of AM than large vocabulary task with a complex language model.

The TIMIT corpus contains recordings of phonetically-balanced prompted English speech. It was recorded using a Sennheiser close-talking microphone at 16 kHz rate with 16 bit sample resolution. TIMIT contains a total of 6300 sentences (5.4 hours), consisting of 10 sentences spoken by each of 630 speakers from 8 major dialect regions of the United States. All sentences were manually segmented at the phone level.

The prompts for the 6300 utterances consist of 2 dialect sentences (SA), 450 phonetically compact sentences (SX) and 1890 phonetically-diverse sentences (SI).

The training set contains 3696 utterances from 462 speakers. The core test set consists of 192 utterances, 8 from each of 24 speakers (2 males and 1 female from each dialect region). The training and test sets do not overlap. 

\subsection{Speech Data, Processing, and Test Description}
As mentioned above, we used TIMIT data available from LDC as a corpus LDC93S1. Then, we ran the Kaldi TIMIT example script s5, which trained various NN-based phone recognition systems with a common HMM-GMM tied-triphone model and alignments. The common baseline system consisted of the following methods: It started from MFCC features which were augmented by $\Delta$ and $\Delta\Delta$ coefficients and then processed by LDA. Final feature vector dimension was 40. We obtained final alignments by HMM-GMM tied-triphone model with 1909 tied-states (may vary slightly if rerun the script). We trained the model with MLLT and SAT methods, and we used fMLLR for the SAT training and a test phase adaptation. We dumped all training, development and test fMLLR processed data, and alignments to disk. Therefore, it was easy to do compatible experiments from the same common starting point. We also dumped MFCC processed by LDA with no normalization and CMN calculated both per speaker and per utterance.

We employed a bigram language/phone model for the final phone recognition. A bigram model is a very weak model for phone recognition; however, it forced focus to the acoustic part of the system, and it boosted benchmark sensitivity. The training, as well as the recognition, was done for 48 phones. We mapped the final results on TIMIT core test set to 39 phones (as is usual for TIMIT corpus processing), and phone error rate (PER) was evaluated by the provided NIST script to be compatible with previously published works. In contrast to the Kaldi recipe, we used a different phone decoder. It is a standard Viterbi-based triphone decoder. It gives better results than the Kaldi standard WFST decoder on the TIMIT phone recognition task.

We have used an open-source Chainer 3.2 DNNs Python tranining tool that supports NVidia GPUs \cite{Chainer}. It is multiplatform and easy to use. 

\subsection{DNN Training}
First, as a reference to RNNs, we trained feed-forward (FF) DNN with ReLU activation function without any pre-training. We used dropout $p=0.2$. We stacked 11 input feature frames to 440 NN input dimension, like in Kaldi example s5.
We have used a network with 8 hidden layers and 2048 ReLU neurons, because it gave the best performance according to our preliminary experiments. The final softmax layer had 1909 neurons. We used SGD with momentum 0.9. The networks were trained in 3 stages with learning rate 1e--2, 4e--3 and 1e--4. The batch size was gradually increased from initial 256 to 1024, and finally to 2048. The training in each stage was stopped when the development data criterion increased in comparison to the last epoch.

Then we have trained LSTM, GRU, reluGRU and M-reluGRU networks. For all of these recurrent networks, we have used identical training setup. We used 4 layers with 1024 units in each. The dropout used was $p=0.2$. We have used output time delay equal to 5 time steps. RNNs were trained in 4 stages. The first stage used Adam optimization algorithm with batch size 512. The other stages used SGD with momentum 0.9, batch size 128 and learning rate equal to 1e--3, 1e--4, and 1e--5 respectivelly. The training in each stage was stopped when the development data criterion increased in comparison to the last epoch, as in FF network case.

We have trained each network on several input data and i-vector combinations.
We used fMLLR data described in the previous section, MFCC and MFCC with CMN. The normalization was calculated either per speaker or per utterance.
For training and testing, we used no i-vectors, online i-vectors and offline i-vectors calculated also either per speaker or per utterance.
We also evaluated online i-vectors for training and offline i-vectors for testing.
The i-vectors were computed according to Kaldi Switchboard example script. However, because of small TIMIT size, we did not use any reduction of data. Entire training dataset was used to estimate i-vector extractor in all steps. The i-vector extractor has been trained only once and online, per-speaker, and per-utterance i-vectors sets were extracted by the same extractor transforms.

Because of stochastic nature of results due to random initialization and stochastic gradient descent, we have performed each experiment 10 times in total.
Then, we have calculated the average phone error rate (PER) and its standard deviation.

\subsection{Results}
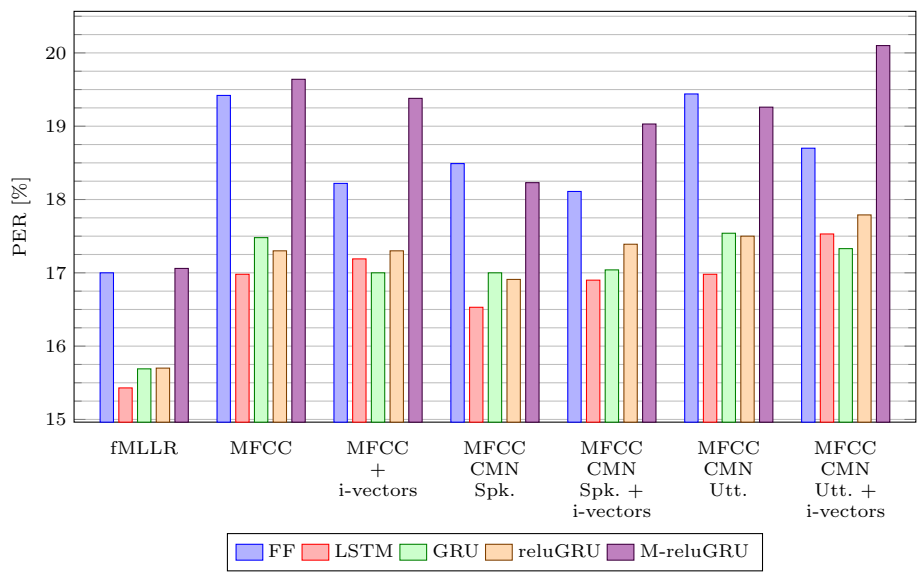
\begin{figure}
    \begin{tikzpicture}
        \scriptsize
        \begin{axis}[
            ybar,
            ylabel={PER [\%]},
            xtick=data,
            xticklabel style={text width=30pt,align=center},
            xtick style={draw=none},
            ymajorgrids=true,
            yminorgrids=true,
            minor y tick num=3,
            bar width=5pt,
            width=360pt,
            height=200pt,
            symbolic x coords={fMLLR,MFCC,MFCC + i-vectors,MFCC CMN Spk.,MFCC CMN Spk. + i-vectors,MFCC CMN Utt.,MFCC CMN Utt. + i-vectors},
            legend style={at={(0.5,-0.27)},anchor=north,legend columns=-1},
            legend image code/.code={
                \draw[#1] (0pt,-3pt) rectangle (10pt,3pt);
            }
        ]
            \addplot table[x=Data, y=FF, col sep=comma] {chart.csv};
            \addplot table[x=Data, y=LSTM, col sep=comma] {chart.csv};
            \addplot[green!50!black,fill=green!20!white] table[x=Data, y=GRU, col sep=comma] {chart.csv};
            \addplot[orange!50!black,fill=orange!30!white] table[x=Data, y=reluGRU, col sep=comma] {chart.csv};
            \addplot[violet!50!black,fill=violet!50!white] table[x=Data, y=M-reluGRU, col sep=comma] {chart.csv};
            \legend{FF,LSTM,GRU,reluGRU,M-reluGRU};
        \end{axis}
    \end{tikzpicture}
    \caption{Phone Error Rate [\%] on Features with Best Performing i-vector Variants}
    \label{fig:chart}
\end{figure}

We have evaluated average PER, its standard deviation for all combinations of three features variants, six i-vector variants, and five NNs architectures. We had to split the results into two tables because of the page size. Table \ref{tbl:per1} shows the average PER for each experiment for FF, LSTM, and GRU NNs architectures. Table \ref{tbl:per2} compares three variants of GRU-based NNs: GRU, reluGRU, M-reluGRU. A subset of the most valuable results is also depicted in Figure \ref{fig:chart}. It is clear that fMLLR adaptation technique worked quite well. All the NN architectures gave the best result with fMLLR. The i-vector adaptation had a stable gain only for FF NN. Two variants of the i-vector adaptation were the best: online i-vectors for training and online or offline per-speaker for testing. Results of RNNs with the i-vector adaptation were interesting, beacause there was no significant gain. The results with adaptation were rather worse. Between RNN architectures, LSTM was the winner (PER 15.43~\% with fMLLR). The GRU and reluGRU gave comparable PERs, 15.7~\% with fMLLR, that was slightly worse than LSTM. M-reluGRU did not performed well and the results were often worse than FF. 

\begin{table}[t]
\centering
\setlength{\tabcolsep}{2.5mm}
\caption{Phone Error Rate [\%] for FF, LSTM and GRU Networks}
\label{tbl:per1}
\begin{tabular}{l|l|l|ccc}
\toprule
& \multicolumn{2}{c|}{i-vectors} & \multicolumn{3}{c}{Phone Error Rate [\%]} \\
\multicolumn{1}{c|}{Data} & \multicolumn{1}{c|}{Training} & \multicolumn{1}{c|}{Testing} & FF & LSTM & GRU \\
\midrule
fMLLR & - & - & $17.00 \pm 0.13$ & $\mathbf{15.43 \pm 0.28}$ & $15.69 \pm 0.19$  \\
 & Off. spk. & Off. spk. & $17.17 \pm 0.16$ & $16.08 \pm 0.19$ & $\mathbf{16.04 \pm 0.29}$  \\
 & Off. utt. & Off. utt. & $17.32 \pm 0.15$ & $\mathbf{16.34 \pm 0.32}$ & $16.43 \pm 0.25$  \\
 & Online & Off. spk. & $17.17 \pm 0.16$ & $\mathbf{16.14 \pm 0.22}$ & $16.15 \pm 0.28$  \\
 & Online & Off. utt. & $17.10 \pm 0.21$ & $16.27 \pm 0.34$ & $\mathbf{16.14 \pm 0.24}$  \\
 & Online & Online & $17.18 \pm 0.14$ & $\mathbf{16.23 \pm 0.26}$ & $\mathbf{16.23 \pm 0.19}$  \\
\midrule
MFCC & - & - & $19.42 \pm 0.18$ & $\mathbf{16.98 \pm 0.27}$ & $17.48 \pm 0.19$  \\
 & Off. spk. & Off. spk. & $19.02 \pm 0.15$ & $\mathbf{17.50 \pm 0.19}$ & $17.63 \pm 0.22$  \\
 & Off. utt. & Off. utt. & $19.29 \pm 0.19$ & $18.12 \pm 0.27$ & $\mathbf{18.09 \pm 0.29}$  \\
 & Online & Off. spk. & $18.22 \pm 0.19$ & $17.19 \pm 0.26$ & $\mathbf{17.00 \pm 0.28}$   \\
 & Online & Off. utt. & $18.48 \pm 0.16$ & $17.27 \pm 0.26$ & $\mathbf{17.21 \pm 0.20}$  \\
 & Online & Online & $18.19 \pm 0.19$ & $\mathbf{17.21 \pm 0.15}$ & $17.33 \pm 0.37$  \\
\midrule
MFCC & - & - & $18.49 \pm 0.19$ & $\mathbf{16.53 \pm 0.20}$ & $17.00 \pm 0.25$  \\
with & Off. spk. & Off. spk. & $18.47 \pm 0.20$ & $\mathbf{17.20 \pm 0.23}$ & $17.33 \pm 0.21$   \\
CMN & Off. utt. & Off. utt. & $18.59 \pm 0.10$ & $17.45 \pm 0.19$ & $\mathbf{17.36 \pm 0.21}$   \\
per & Online & Off. spk. & $18.11 \pm 0.24$ & $\mathbf{16.90 \pm 0.24}$ & $17.04 \pm 0.16$   \\
speaker & Online & Off. utt. & $18.17 \pm 0.22$ & $17.34 \pm 0.31$ & $\mathbf{17.06 \pm 0.20}$   \\
 & Online & Online & $18.21 \pm 0.19$ & $17.25 \pm 0.26$ & $\mathbf{17.24 \pm 0.31}$  \\
\midrule
MFCC & - & - & $19.44 \pm 0.27$ & $\mathbf{16.98 \pm 0.20}$ & $17.54 \pm 0.20$  \\
with & Off. spk. & Off. spk. & $19.10 \pm 0.17$ & $\mathbf{17.60 \pm 0.31}$ & $17.64 \pm 0.33$   \\
CMN & Off. utt. & Off. utt. & $19.32 \pm 0.14$ & $18.28 \pm 0.35$ & $\mathbf{18.15 \pm 0.35}$   \\
per & Online & Off. spk. & $18.70 \pm 0.18$ & $17.53 \pm 0.23$ & $\mathbf{17.33 \pm 0.18}$   \\
utterance & Online & Off. utt. & $18.63 \pm 0.16$ & $17.60 \pm 0.23$ & $\mathbf{17.46 \pm 0.19}$   \\
 & Online & Online & $18.73 \pm 0.18$ & $17.66 \pm 0.23$ & $\mathbf{17.43 \pm 0.19}$   \\
\bottomrule
\end{tabular}
\end{table}

\begin{table}[t]
\centering
\setlength{\tabcolsep}{2.5mm}
\caption{Phone Error Rate [\%] for GRU and Its Modifications}
\label{tbl:per2}
\begin{tabular}{l|l|l|ccc}
\toprule
& \multicolumn{2}{c|}{i-vectors} & \multicolumn{3}{c}{Phone Error Rate [\%]} \\
\multicolumn{1}{c|}{Data} & \multicolumn{1}{c|}{Training} & \multicolumn{1}{c|}{Testing} & GRU & reluGRU & M-reluGRU \\
\midrule
fMLLR & - & - & $\mathbf{15.69 \pm 0.19}$ & $15.70 \pm 0.56$ & $17.06 \pm 0.77$ \\
 & Off. spk. & Off. spk. & $\mathbf{16.04 \pm 0.29}$ & $16.28 \pm 0.38$ & $17.50 \pm 0.72$ \\
 & Off. utt. & Off. utt. & $16.43 \pm 0.25$ & $\mathbf{16.33 \pm 0.13}$ & $18.25 \pm 0.85$ \\
 & Online & Off. spk. & $\mathbf{16.15 \pm 0.28}$ & $16.19 \pm 0.22$ & $17.76 \pm 0.94$ \\
 & Online & Off. utt. & $\mathbf{16.14 \pm 0.24}$ & $16.23 \pm 0.18$ & $17.85 \pm 0.76$ \\
 & Online & Online & $\mathbf{16.23 \pm 0.19}$ & $16.39 \pm 0.33$ & $17.60 \pm 0.67$ \\
\midrule
MFCC & - & - & $17.48 \pm 0.19$ & $\mathbf{17.30 \pm 0.50}$ & $19.64 \pm 1.05$ \\
 & Off. spk. & Off. spk. & $\mathbf{17.63 \pm 0.22}$ & $18.32 \pm 0.39$ & $20.13 \pm 0.93$ \\
 & Off. utt. & Off. utt. & $\mathbf{18.09 \pm 0.29}$ & $18.35 \pm 0.37$ & $20.70 \pm 0.65$ \\
 & Online & Off. spk. & $\mathbf{17.00 \pm 0.28}$ & $17.30 \pm 0.38$ & $19.38 \pm 0.96$ \\
 & Online & Off. utt. & $\mathbf{17.21 \pm 0.20}$ & $17.52 \pm 0.47$ & $19.44 \pm 0.89$ \\
 & Online & Online & $\mathbf{17.33 \pm 0.37}$ & $17.41 \pm 0.44$ & $19.29 \pm 0.89$ \\
\midrule
MFCC & - & - & $17.00 \pm 0.25$ & $\mathbf{16.91 \pm 0.22}$ & $18.23 \pm 0.53$ \\
with & Off. spk. & Off. spk. & $\mathbf{17.33 \pm 0.21}$ & $17.70 \pm 0.39$ & $19.44 \pm 0.66$ \\
CMN & Off. utt. & Off. utt. & $\mathbf{17.36 \pm 0.21}$ & $17.91 \pm 0.35$ & $19.43 \pm 1.17$ \\
per & Online & Off. spk. & $\mathbf{17.04 \pm 0.16}$ & $17.39 \pm 0.27$ & $19.03 \pm 1.07$ \\
speaker & Online & Off. utt. & $\mathbf{17.06 \pm 0.20}$ & $17.48 \pm 0.29$ & $18.93 \pm 0.74$ \\
 & Online & Online & $\mathbf{17.24 \pm 0.31}$ & $17.45 \pm 0.27$ & $18.89 \pm 0.78$ \\
\midrule
MFCC & - & - & $17.54 \pm 0.20$ & $\mathbf{17.50 \pm 0.29}$ & $19.26 \pm 0.85$ \\
with & Off. spk. & Off. spk. & $\mathbf{17.64 \pm 0.33}$ & $18.05 \pm 0.27$ & $19.08 \pm 0.77$ \\
CMN & Off. utt. & Off. utt. &$\mathbf{18.15 \pm 0.35}$ & $18.52 \pm 0.33$ & $21.04 \pm 0.97$ \\
per & Online & Off. spk. & $\mathbf{17.33 \pm 0.18}$ & $17.79 \pm 0.31$ & $20.10 \pm 0.95$ \\
utterance & Online & Off. utt. & $\mathbf{17.46 \pm 0.19}$ & $18.05 \pm 0.24$ & $19.63 \pm 0.99$ \\
 & Online & Online & $\mathbf{17.43 \pm 0.19}$ & $17.85 \pm 0.18$ & $20.01 \pm 0.69$ \\
\bottomrule
\end{tabular}
\end{table}

    \section{Conclusion}
    In this paper, we have compared feed-forward and several recurrent network architectures on input data with fMLLR or i-vector adaptation techniques.
The used recurrent networks were based on LSTM and GRU units.
We have also evaluated two GRU modifications: reluGRU, with ReLU activation function, and M-reluGRU, with ReLU activation function and without the reset gate.
As features, we have used MFCC processed by LDA without normalization or with CMN calculated either per speaker or per utterance, and also fMLLR adaptation.
We have also augmented the features with several variants of i-vectors: online or offline calculated either per speaker or per utterance.
Due to the stochastic nature of the used optimizers, we have performed all experiments 10 times in total and calculated the average phone error rate and its standard deviation.

For all networks, we have obtained the best results with fMLLR adaptation.
The i-vector adaptation consistently improved the results only for FF networks.
In the case of RNN, i-vectors did not lead to any significant improvement; it even gave worse results in all LSTM experiments and in some experiments with GRU variants.
We have achieved the best results with LSTM network (PER 15.43~\% with fMLLR).
GRU and reluGRU were slightly worse (both having PER 15.7~\% with fMLLR).
M-reluGRU was in some cases even worse than FF network.

For all our experiments, we have used Chainer 3.2 DNN training framework with Python programming language and we have published our open-source scripts at \url{https://github.com/OrcusCZ/NNAcousticModeling} to easily replicate the results and to help continue the development.

    \section*{Acknowledgement}
    This work was supported by the project no. P103/12/G084 of the Grant Agency of the Czech Republic and by the grant of the University of West Bohemia, project No. SGS-2016-039.
    Access to computing and storage facilities owned by parties and projects contributing to the National Grid Infrastructure MetaCentrum provided under the programme "Projects of Large Research, Development, and Innovations Infrastructures" (CESNET LM2015042), is greatly appreciated.

    \newpage
    %
    %
    %
    \bibliographystyle{splncs04}
    \bibliography{SLSP2018_Vanek_Michalek}
\end{document}